\documentclass[10pt]{article}
\usepackage{graphics}
\usepackage{conf_iap}
\begin{document}
\newcommand{\apjl}{ApJL\,\,}
\newcommand{\aap}{A\&A\,\,}
\newcommand{\aaps}{A\&AS\,\,}
\newcommand{\mnras}{MNRAS\,\,}
\vglue -2cm 
\small{\noindent To appear in XVIIIth IAP Colloquium,
\emph{On the Nature of Dark Energy:
observational and theoretical results on the accelerating universe}, ed. 
P. Brax, J. Martin \& J. P. Uzan}

\vskip 1.5cm

\heading{Constraints on dark energy and quintessence\\ 
with a comoving standard ruler applied to 2dF quasars} 
\par\medskip\noindent
\author{Gary A. Mamon$^{1,2}$, Boud F. Roukema$^{3}$}
\address{Institut d'Astrophysique de Paris, 98 bis Bd Arago, 
F--75014 Paris, FRANCE}
\address{GEPI, Observatoire de Paris, 61 av de l'Observatoire, F--75014 Paris,
FRANCE}
\address{Torun Centre for Cosmology,
University of Nicolas Copernicus,
ul. Gagarina 11,
PL--87100 Torun, POLAND}

\begin{abstract}
Structures on very large scales ($> 100 \,\rm Mpc$) have negligible peculiar
motions, and are thus roughly fixed in comoving space.
We looked for significant
peaks at very large separation 
in the two-point correlation function --- corrected for redshift
selection effects --- of
a well convered subsample of 2378 quasars of the
recently released 10k sample of the {\sf 2dF} quasar survey.
Dividing our sample in three redshift intervals, we find a peak at
$\simeq 244 \, h^{-1} \, \rm Mpc$, which is perfectly comoving for a
restricted set of cosmological parameters, namely $\Omega_m = 0.25\pm0.15$
and $\Omega_\Lambda=0.65\pm0.35$ (both at 95\% confidence).
Assuming a flat Universe, we constrain the quintessence parameter $w_Q <
-0.35$ (95\% confidence). We discuss the compatibility of our analysis with
possible peaks in the power spectrum.
\end{abstract}

\section{Introduction}
The quest for the parameters of the primordial Universe has significantly
advanced in the last 10 years.
There is a general agreement that $\Omega_m \simeq 0.3$, for
example from
the internal kinematics of clusters 
and groups 
\cite{M93_aussois_dyn}, from the gas fraction in clusters
\cite{WNEF93} and groups \cite{HM94}.
Supernovae used as standard candles have led to
$\Omega_\Lambda > 0$, with a degeneracy with $\Omega_m$. Finally, the scales
of the angular fluctuations of the cosmic microwave background (CMB) lead
to a nearly perfectly flat Universe $\Omega_m\!+\!\Omega_\Lambda = 1$.
The combination of constraints from supernovae
and the CMB recovers $\Omega_m \simeq 0.3$ (with $\Omega_\Lambda \simeq 0.3$).

We present below a measurement of the cosmological parameters, with a fairly
new technique based upon using very large scale structures ($> 100\,\rm Mpc$
in size) as comoving standard rulers.

\section{Very large-scale structures should follow the Hubble flow}

Very large-scale structures
should follow the Hubble
flow.
The rms peculiar velocity within a ball of radius $R$ is given by
\begin{equation}
v_{\rm rms}^2 (R) = {1\over 2\,\pi^2} \,(H\,f)^2\,\int_0^\infty \widetilde
W^2(k\,R)\,P(k)\,dk \ ,
\label{vrms}
\end{equation}
where for $\Omega_\Lambda = 0$ or for flat Universes, $f \simeq
\Omega_m^{0.6}$, while $P(k)$ is the primordial density fluctuation
spectrum,\footnote{We adopt, as the observers, 
the high definition of $P(k)$ ($[2\,\pi]^3$ higher than the low one).} 
and $\widetilde W(x) = 3\,(\sin x - x\,\cos x)/x^3$ is the Fourier transform
of the top-hat smoothing kernel.
Eq.~(\ref{vrms}) leads to a present-day value
$v_{\rm rms,0} < 330 \, \rm km \, s^{-1}$ for $R > 100 \, h^{-1} \,
\rm Mpc$.
Since the peculiar velocity increases with time, one simply finds that the
peculiar motion satisfies
$\delta R < v_{\rm rms,0} \,t_0 = 3.3 \, h^{-1} \, \rm Mpc$,
so that peculiar motions account for less than 3\% for structures larger than
$100 \, h^{-1} \, \rm Mpc$.
Thus, VLSS follow the Hubble expansion to good precision,
or, in other words,
\emph{very large scale structures should be comoving standard rulers}.

\section{Is their a specific scale for very large scale structures?}

Since the report \cite{BEKS90} of a redshift 
periodicity of structures on the scale
of $130 \, h^{-1} \, {\rm Mpc} = 2\,\pi/(0.048\,h\,\rm Mpc^{-1})$, 
the existence of a preferred scale for very
large scale structures, has been a matter of debate.
\begin{table}[ht]
\begin{center}
\caption{Power spectra from wide surveys}
{\small
\begin{tabular}{rcccrllc}
\multicolumn{1}{l}{Ref.} & Year & Survey & Objects & \multicolumn{1}{c}{Number} & \multicolumn{1}{c}{$\!\!\!\!$Peak} & \multicolumn{1}{c}{$\!\!\!\!$Dip} & Notes \\
\cline{6-7}
	&	&	&	&	& \multicolumn{2}{c}{($h\,\rm Mpc^{-1}$)} & \\
\hline
\cite{Einasto+97} & 1997 & ACO & clusters & 869 & 0.05 & \\
\cite{TED98} & 1998 & APM & clusters & 364 & & 0.08 & \\
\cite{MB01} & 2001 & ACO & clusters & 637 & & 0.04 \\
\cite{Schuecker+01} & 2001 & REFLEX & clusters & 452 & & \\
\cite{ZAL01} & 2001 & XBACS & clusters & 242 & & 0.035 \\
\cite{ZAL01} & 2001 & BCS & clusters & 301 & 0.08 & \\
\hline
\cite{Lin+96} & 1996 & LCRS & galaxies & 24000 & & \\ 
\cite{HT02} & 2000 & PSCz & galaxies & 12400 & 0.028 & 0.036 & 1 \\
\cite{Percival+01} & 2001 & 2dF & galaxies & 100000 & & 0.1 & 2 \\
\cite{THX02} & 2002 & 2dF & galaxies & 100000 & 0.05 & 0.1 & 3 \\
\hline
\cite{Hoyle+02} & 2002 & 2dF & quasars & 10000 & 0.069 & & \\
\hline
\end{tabular}
}
\end{center}
{\small 
Notes: 1) marginal peak and dip; 2) not deconvolved for survey geometry; 3) 
$(\Omega_m,\Omega_\Lambda) = (0.2,0.8)$}
\label{surveys}
\end{table}
As seen in Table 1, whereas many authors find significant
peaks and dips in $P(k)$, measured from large surveys,
they do not agree on their positions.
In what follows, we will assume that there \emph{is} a feature (peak or dip)
in $P(k)$, which
should lead to a quasi-periodic signal in the space
distribution that is essentially frozen in comoving space.

\section{The comoving standard ruler applied to quasars}

After previous attempts \cite{RM00,RM01} on a low spectral resolution survey
of quasars \cite{ICS96}, we have analyzed the 2dF quasar sample, called {\sf
2QZ-10k} \cite{Croom+01}, publicly
made available in the Spring of 2001.
After discarding poorly sampled regions (where the survey completeness was
less than 80\%), leaving us with 2378 quasars, 
we computed the spatial two-point correlation function of
the quasars in 3 redshift bins ($z = [0.6,1.1], [1.1, 1.6], [1.6, 2.1]$), 
looking for features at the same
comoving separations. 
The correlation functions were estimated as \cite{LS93}
$\xi(r) = [\hbox{DD} - 2\,\hbox{DR}/n + \hbox{RR}/n^2]/(\hbox{RR}/n^2)$,
where DD, DR and RR indicate the number of data-data, data-random, and
random-random quasar pairs respectively.
We avoided redshift selection effects (emission lines will be visible in some
intervals of redshift), by constructing our random catalogs with random
angular positions, but redshifts obtained by scrambling the distribution in
the dataset.
The correlation functions were smoothed with a $15 \, h^{-1} \, \rm Mpc$
gaussian.

We then repeated the exercise for $21\!\times\!21$ pairs of
$(\Omega_m,\Omega_\Lambda)$, since the separation of two quasars depends on
their angular separation and their two redshifts in a non-trivial function of
$\Omega_m$ and $\Omega_\Lambda$.

Only in the region $\Omega_\Lambda = 1.4\,\Omega + 0.15 \pm 0.35$ did we
obtain strong peaks in the correlation functions.
Figure~1 \cite{RMB02} illustrates the subregion --- the vertical ellipse
centered on 
$\Omega_m=0.25\pm0.15$ and $\Omega_\Lambda=0.65\pm0.35$ (95\% confidence) ---
where
the peaks (at separation $r = 244 \, h^{-1} \, \rm Mpc$)
in the correlation function are at the same comoving separation in the
three redshift bins.
\emph{This appears to be the strongest joint constraint on $\Omega_m$ and
$\Omega_\Lambda$ from a single survey.}
\begin{figure}[ht]
\begin{center}
{\resizebox{!}{0.49\hsize}{\includegraphics{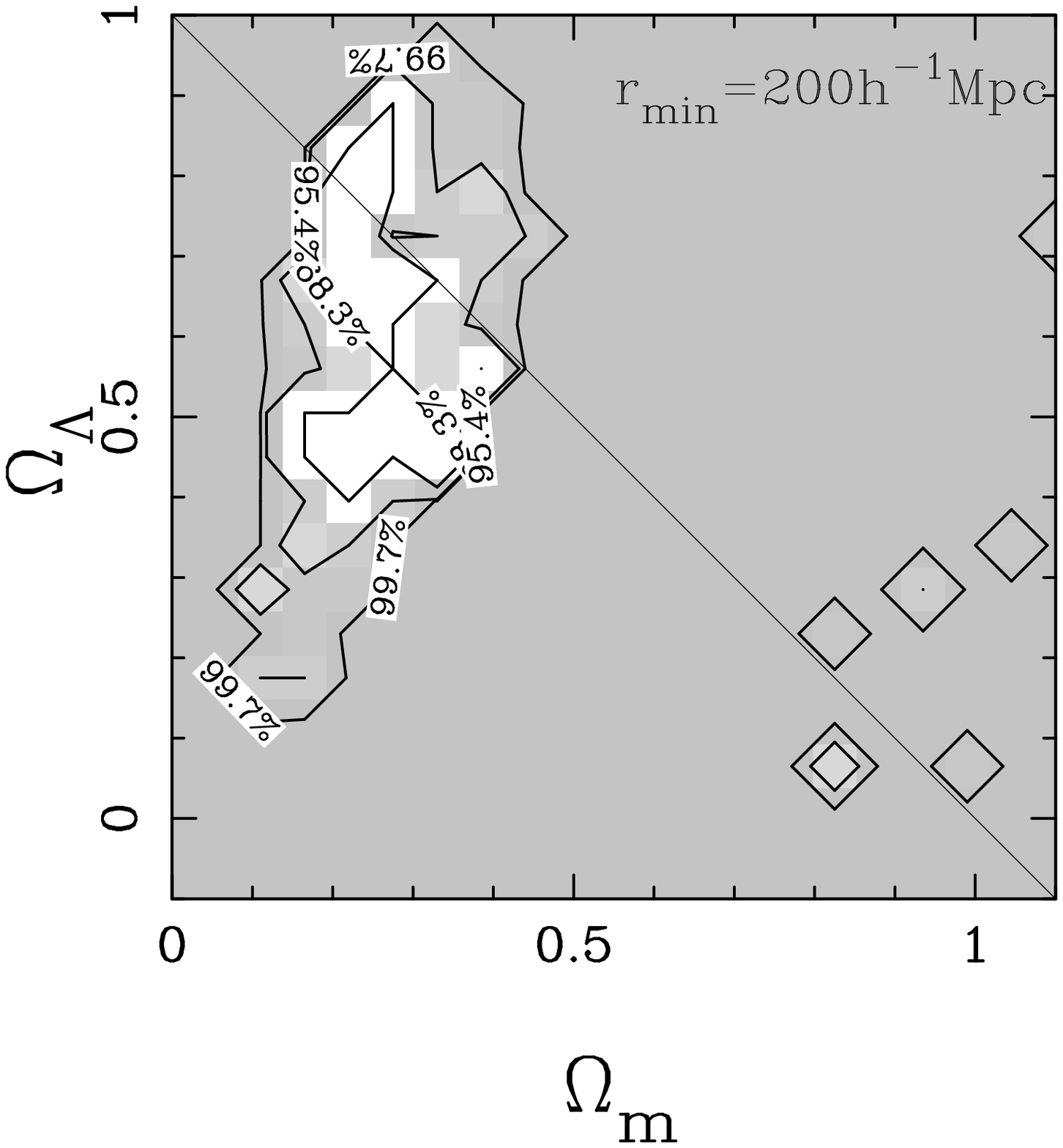}}}
{\resizebox{!}{0.49\hsize}{\includegraphics{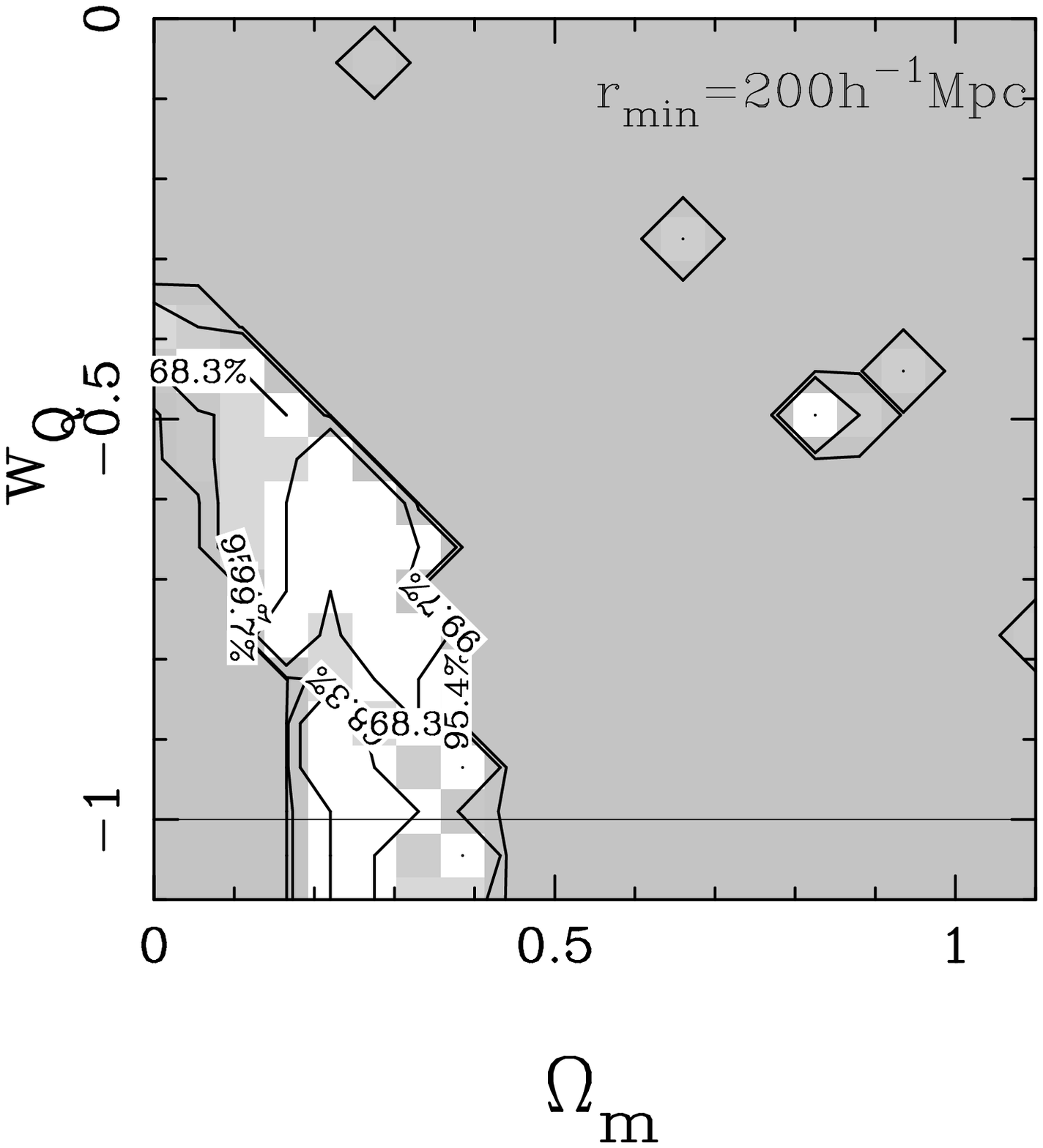}}}
\caption[]{\emph{left}: $\Omega_m$ and $\Omega_\Lambda$ pairs for significant
comoving peaks in the correlation function of {\sf 2QZ-10k} quasars.
\emph{right}: 
Same for $\Omega_m$ and $w_Q$ (quintessence) pairs, assuming a
flat Universe. From \cite{RMB02}.}
\end{center}
\end{figure}

Figure 1 also displays the constraints on quintessence
for the flat case ($w_Q =-1$ in the standard model), and
we find $w_Q < -0.35$ (95\% confidence).

As a caveat, we have computed the expected correlation function 
from the power spectrum of \cite{Hoyle+02}, which we have padded with a
CMBFAST calculation with $(\Omega_m,\Omega_\Lambda,\Omega_b,h,\sigma_8) =
(0.3,0.7,0.1,0.6,0.9)$ 
that fits well their $P(k)$, using the relation
\begin{equation}
\xi(r) = {1\over 2\pi^2}\,\int_0^\infty k^2\, P(k)\,{\sin(kr) \over kr}\,dk \ .
\label{xiofr}
\end{equation}
We find oscillations in the $15 \, h^{-1} \, \rm Mpc$ gaussian smoothed
correlation function that are 25 times smaller than found by \cite{RMB02}.
We check this by considering a feature in $P(k)$ at wavenumber $k_0$
that has amplitude $L^3$ and
width $\Delta$ decades in $\log k$.
If $\Delta \ll 1$, Eq.~(\ref{xiofr}) yields
$\xi \simeq \ln10/(2\pi^2)\,(k_0\,L)^3 \Delta \allowbreak
\sin(k_0 r) / (k_0 r)$.
If $\xi(r) \propto \sin(k_0 \,r)/(k_0\,r)$ has a secondary maximum at
$k_0\,r_0 = 5\pi/2$, then
$L^3\,\Delta = 8\,\xi(r_0) \,r_0^3/( 25\,\ln 10)$.
For $r_0 = 244 \, h^{-1} \, \rm Mpc$ and
$\xi(r_0) \simeq 0.03$ \cite{RMB02}, we need $L^3 \,\Delta\simeq 6\times10^4\,
h^{-3} \, {\rm Mpc}^3 = 1.5\,P(k_0)$. So if the spike is, say, one-tenth of a
decade wide, we need to locally boost $P(k)$ by a factor 15, whereas the
measured peak \cite{Hoyle+02} is only a factor 2 above the interpolated
$P(k)$.

It seems clear that there is an inconsistency between these two
analyses, and we
are presently redoing our analysis of $\xi(r)$ from {\sf 2QZ-10k} 
to resolve this question. Whether
$\xi(r)$ or $P(k)$ is the better tool for detecting comoving
standard rulers across distinct redshift intervals remains an open question.

\acknowledgements{We thank Jim Fry for helpful discussions on power spectra.}

\vfill
\end{document}